# Terahertz Metamaterials for Linear Polarization Conversion and Anomalous Refraction

Nathaniel K. Grady[1], Jane E. Heyes[1], Dibakar Roy Chowdhury[1], Yong Zeng[2], Matthew T. Reiten[1], Abul K. Azad[1], Antoinette J. Taylor[1], Diego A. R. Dalvit[2], and Hou-Tong Chen[1]*

[1]Center for Integrated Nanotechnologies, MS K771, Los Alamos National Laboratory, Los Alamos, New Mexico 87545, USA.

[2]Theoretical Division, MS B213, Los Alamos National Laboratory, Los Alamos, New Mexico 87545, USA.

*Correspondence to:  chenht@lanl.gov.

**Polarization is one of the basic properties of electromagnetic waves conveying valuable information in signal transmission and sensitive measurements. Conventional methods for advanced polarization control impose demanding requirements on material properties and attain only limited performance. Here, we demonstrate ultrathin, broadband, and highly efficient metamaterial-based terahertz polarization converters that are capable of rotating a linear polarization state into its orthogonal one. Based on these results we create metamaterial structures capable of realizing near-perfect anomalous refraction. Our work opens new opportunities for creating high performance photonic devices and enables emergent**

**metamaterial functionalities for applications in the technologically difficult terahertz frequency regime.**

Control and manipulation of electromagnetic (EM) polarization states has greatly impacted our daily life from consumer products to high-tech applications. Conventional state-of-the-art polarization converters utilize birefringence or total internal reflection effects (*1*) in crystals and polymers, which cause phase retardation between the two orthogonally polarized wave components. Expanding their typically limited bandwidth requires complex designs using multilayered films or Fresnel rhombs. At microwave and millimeter wave frequencies, narrowband polarization converters have been constructed using metallic structures, such as birefringent multilayered meander-line gratings (*2*). Fabrication challenges and high losses render these unsuitable for optical frequencies (*3*).

Metamaterials have enabled the realization of many phenomena and functionalities unavailable using naturally occurring materials (*4-6*). Many basic metamaterial structures, such as metal split-ring resonators (*7*), exhibit birefringence suitable for polarization conversion (*8-16*), which has been mostly investigated in the microwave frequency range. Broadband metamaterial circular polarizers have been demonstrated in the optical regime using gold helix structures (*17*) and stacked nano-rod arrays with a tailored rotational twist (*18*). Metamaterial-based polarimetric devices are particularly attractive in the terahertz (THz) frequency range due to the lack of suitable natural materials for THz device applications. However, the currently available designs suffer from either very limited bandwidth or high losses (*19-21*). In this work, we demonstrate high-efficiency and broadband linear THz polarization conversion using ultrathin planar metamaterials. In addition, our designs enable a dramatic improvement of the recently demonstrated anomalous (or generalized laws of) reflection/refraction (*22, 23*) by eliminating the ordinary components.

Our first metamaterial linear polarization converter design (Fig. 1A, B) operates in reflection and consists of a metal cut-wire array and a metal ground plane separated by a dielectric spacer. We consider an incident wave $\boldsymbol{E}_0$ linearly polarized in the *x*-direction. It excites a dipolar oscillation $\boldsymbol{p}$ mainly along the cut-wires, which has parallel ($p_x$) and perpendicular ($p_y$) components to $\boldsymbol{E}_0$. While $\boldsymbol{E}_0$ and $p_x$ determine the co-polarized scattered field, $p_y$ results in cross-polarized scattering, forming the dispersive reflection and transmission (Fig. S1) of the cut-wire array. Without the ground plane, the polarization conversion efficiency is low and the overall reflection and transmission are elliptically polarized. The ground plane and the cut-wire array form a Fabry-Pérot-like cavity (*24, 25*); the consequent interference of polarization couplings in the multireflection process may either enhance or reduce the overall reflected fields with co- and cross-polarizations (for more details, see Supplementary Materials (*26*)).

We validate this concept by performing full-wave numerical simulations shown in Fig. 1C (for the polarization angle dependence, see Fig. S2). Between 0.7 and 1.9 THz, the cross-polarized reflection carries more than 50% of the incident power, and the co-polarized component is mostly below 20%. Between 0.8 and 1.36 THz, the cross-polarized reflection is higher than 80% and the co-polarized one below 5%, representing a broadband and high-performance linear polarization converter in reflection. Further numerical simulations reveal that this broadband and high efficiency performance is sustained over a wide incidence angle range (Fig. S3). The broadband operation results from the superposition of multiple polarization conversion peaks around 0.8 THz, 1.2 THz, and 1.9 THz in Fig. 1C, where the efficiency is mainly limited by dielectric loss. At other frequencies the co-polarized reflection also contributes.

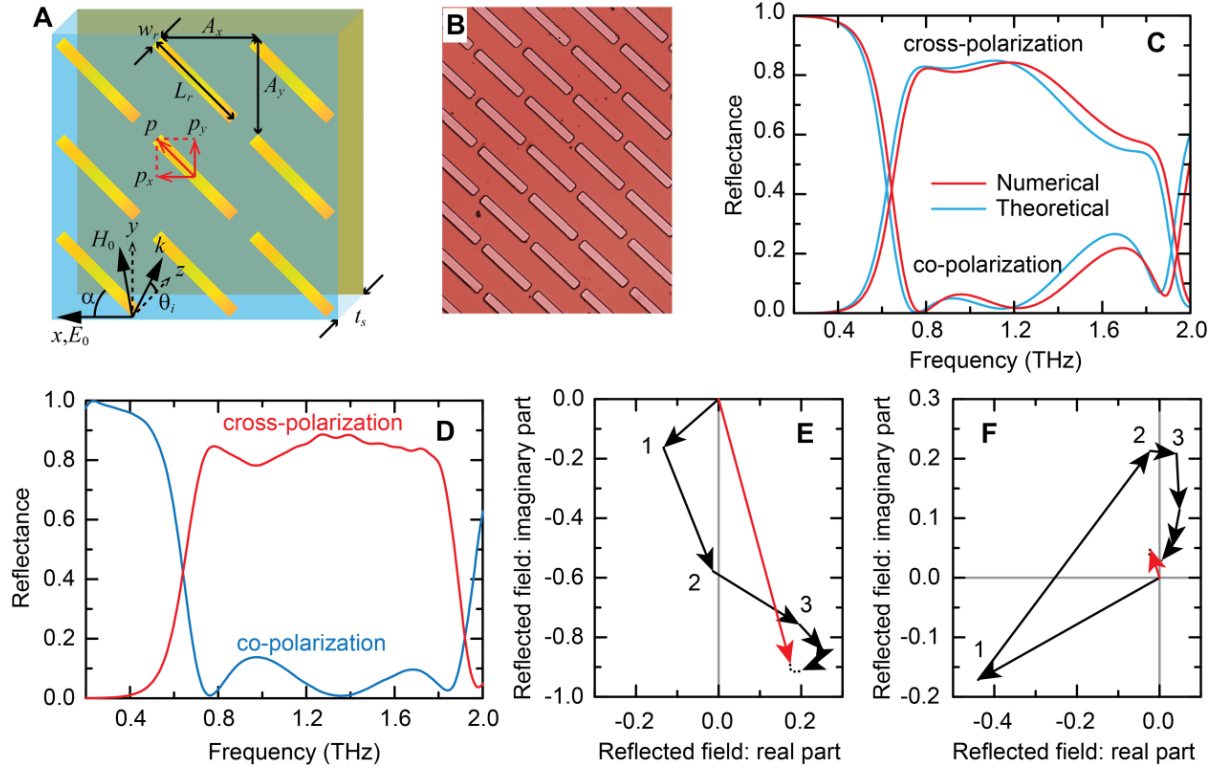

**Fig. 1. Broadband polarization conversion in reflection.** (**A**) Schematic and (**B**) optical micrograph of the metamaterial linear polarization converter. Both the gold cut-wire array and the gold ground plane are 200 nm thick, and they are separated by a polyimide dielectric spacer with thickness $t_s$ = 33 μm and dielectric constant $\varepsilon = 3(1 + 0.05i)$. The periodicity $A_x = A_y$ = 68 μm, cut-wire length $L_r$ = 82 μm, and width $w_r$ = 10 μm. The incidence angle $\theta_i = 25°$, and the incident electric field $\boldsymbol{E}_0$ is linearly polarized in the *x*-direction (i.e. *s*-polarized), with an angle $\alpha = 45°$ with respect to the cut-wire orientation. (**C**) Numerically simulated and theoretically calculated, and (**D**) experimentally measured co- and cross-polarized reflectance. (**E**) Cross- and (**F**) co-polarized multiple reflections theoretically calculated at 0.76 THz, revealing the constructive and destructive interferences, respectively. Similar behavior occurs for other frequencies as well. The numbers *j* (1, 2, 3) indicate the (*j*–1)-th roundtrip within the device. The red arrows are the converged cross- and co-polarized reflected fields.

The underlying reason for the enhanced polarization conversion is the interference between the multiple polarization couplings in the Fabry-Pérot-like cavity. To confirm this interpretation, we calculate the co- and cross-polarized reflected fields caused by each roundtrip within the cavity (Fig. S5). For each individual reflection we plot the complex reflected field in Figs. 1E, F. As expected, the superposition of these partial cross- (co-) polarized reflected fields results in a constructive (destructive) interference and gives a nearly unity (zero) overall cross- (co-) reflection (denoted by the red arrows in Figs. 1E, F). The calculated overall reflections (Fig. 1C) are in excellent agreement with both the numerical simulations and experimental data. See Supplementary Materials for more details (*26*).

Many applications require linear polarization conversion in transmission mode, for which the metal ground plane must be replaced. Our solution is to use a metal grating which transmits the cross-polarized waves while still acting as a ground plane for co-polarized waves. To retain the backward propagating cross-polarized waves without blocking the incident waves, we add an orthogonal metal grating in front of the cut-wire (Fig. 2A). In addition, we add a 4 μm thick polyimide capping layer both before the front and behind the back grating. This ultrathin, freestanding device serves as a high performance linear polarization converter in transmission with reduced co- and cross-polarized reflections. In Fig. 2B we plot the numerically simulated cross-polarized transmittance and co-polarized reflectance, for normal incidence with an *x*-polarized incident electric field (this device also operates over a wide incidence angle range (Fig. S6)). Also shown in Fig. 2B are the theoretical multireflection model results (*26*) and measured data for our fabricated device. There is excellent agreement among the numerical, experimental, and theoretical results. The device is able to rotate the linear polarization by 90°, with a conversion efficiency exceeding 50% from 0.52 to 1.82 THz, with the highest efficiency of 80% at 1.04 THz. We emphasize that, in both simulation and experiments, the co-polarized transmission and cross-polarized reflection (not shown) are practically zero due to the use of gratings. The measured ratio between the co- and cross-polarized transmittance is less than 0.1 between 0.2 and 2.2 THz, covering the whole frequency range in a typical THz-TDS. The device performance, limited by the dielectric loss and co-polarized reflection (Fig. 2B), can be further improved through optimizing the structural design and using lower loss dielectric materials.

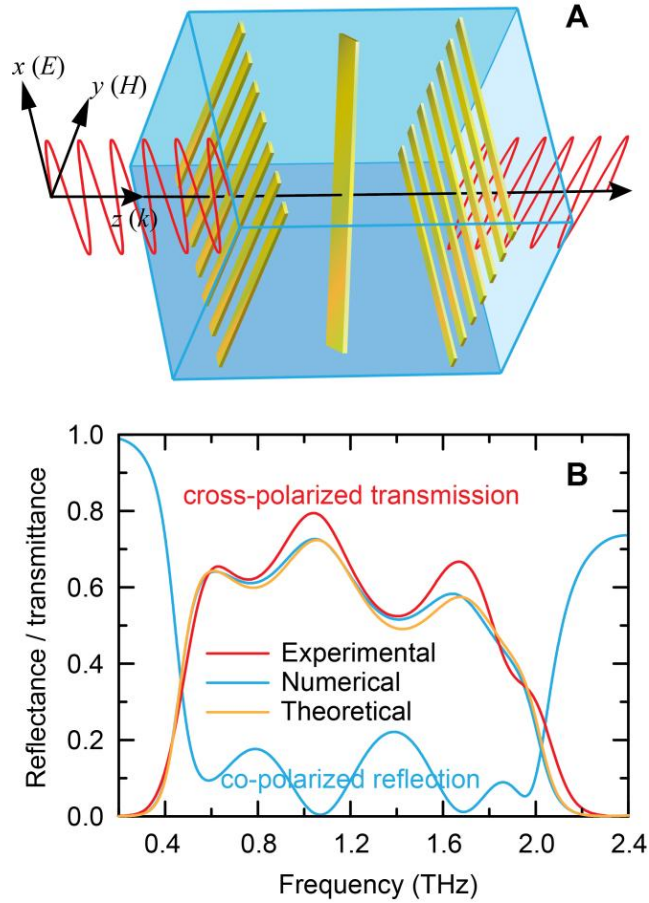

**Fig. 2. Broadband polarization conversion in transmission. (A)** Schematic of the unit cell of the metamaterial linear polarization converter, where a normally incident *x*-polarized wave is converted into a *y*-polarized one. The cut-wire array is the same as in Fig. 1, the spacer is polyimide, and the separation between the cut-wire array and the gratings is 33 μm. The gold grating wire width is 4 μm, periodicity is 10 μm, and thickness is 200 nm. For this freestanding device, the gratings are covered with 4 μm thick polyimide caps. **(B)** Cross-polarized transmittance obtained through experimental measurements, numerical simulations, and theoretical calculations. Also shown is the numerically simulated co-polarized reflectance.

Recent demonstrations of the general laws of reflection/refraction and wavefront shaping (*22, 23*) rely on creating a phase gradient in the cross-polarized scattering from anisotropic metamaterials. However, the single-layered metamaterial only produced weak anomalously reflected/refracted beams with most of the power remaining in the ordinary beams. Our demonstration of high-efficiency linear polarization converters allows us to accomplish broadband near-perfect anomalous reflection/refraction by largely eliminating the ordinary beams. We use eight anisotropic resonators with various geometries and dimensions in a super-unit-cell (Figs. 3A, B) to create a linear phase variation of the cross-polarized transmission spanning a $2\pi$ range. The resonator dimensions are determined by numerical simulations and specified in Fig. S7. Each of these resonators can be used to construct a high-performance linear polarization converter with similar cross-polarized transmission but a phase increment of approximately $\pi/4$ (Fig. 3C). Therefore, when combined into the super-unit-cell shown in Figs. 3A, B we expect a linear phase gradient of the cross-polarized transmitted wavefront, resulting in anomalous refraction.

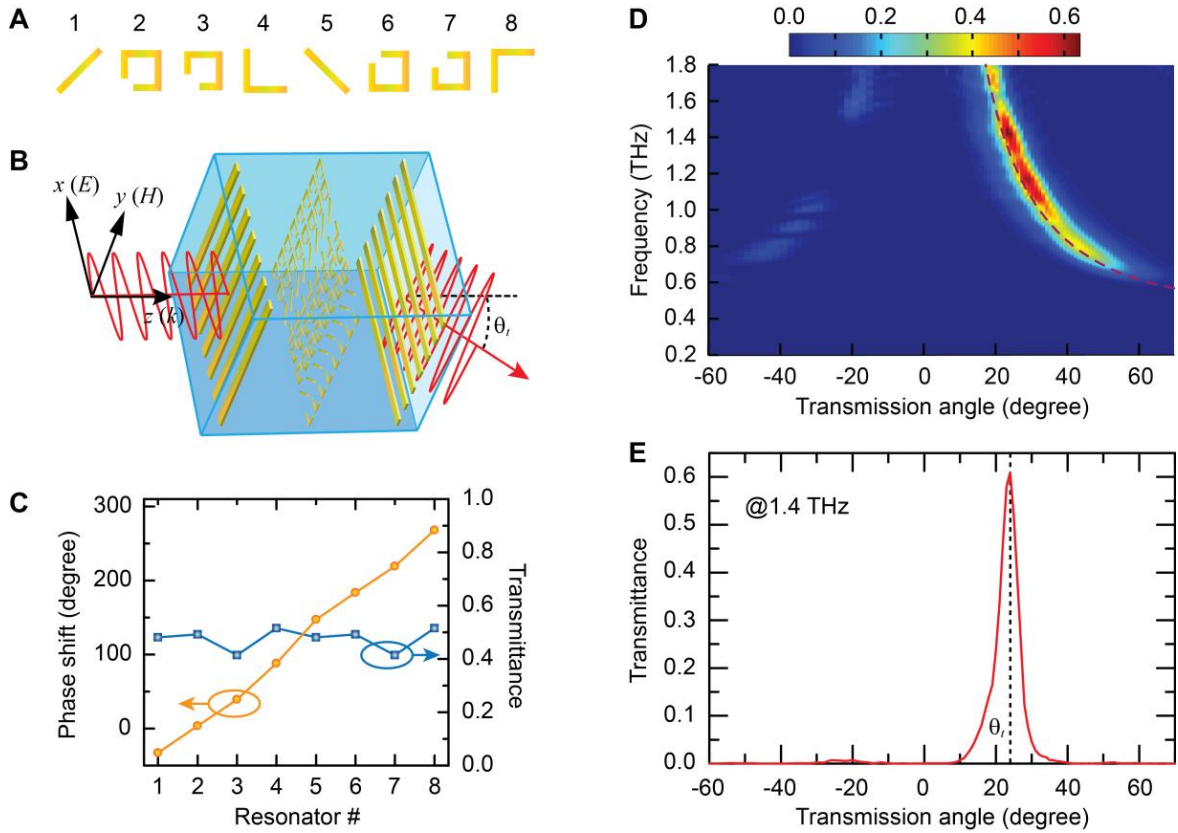

**Fig. 3. Broadband and near-perfect anomalous refraction.** (**A**) Resonator array super-unit-cell within the (**B**) anomalous refraction design (not to scale). A normally incident *x*-polarized wave is converted to a *y*-polarized transmission beam, which bends in the *x-z* plane to an angle $\theta_t$ with respect to the *z*-axis. (**C**) The simulated cross-polarized transmittance and corresponding phase shift when each individual resonator is used in the linear polarization converter. (**D**) Experimentally measured cross-polarized transmittance as a function of frequency and angle. The dashed curve is the theoretically calculated frequency-dependent bending angle. (**E**) Cross-polarized transmittance at 1.4 THz under normal incidence as a function of angle.

We characterize our fabricated free-standing sample under normal incidence ($\theta_i = 0$) (*26*). In Fig. 3D we plot the cross-polarized transmittance as a function of frequency and transmission angle, while the measured co-polarized transmission is negligible (not shown). Over a broad bandwidth, the ordinary refraction ($\theta_t = 0$) is practically zero, and only the anomalous beam is transmitted at a frequency-dependent refraction angle $\theta_t$ following the generalized law of refraction (*22*): $n_2\sin(\theta_t) - n_1\sin(\theta_i) = d\Phi/dx$, where $n_1 = n_2 = 1$ for the surrounding air, $d\Phi/dx = \lambda_0/D_1$ is the phase gradient imposed by the sample, $\lambda_0$ is the free space wavelength, and $D_1 = 560$ µm is the periodicity of the super-unit-cell along the phase gradient (*x*-direction). The calculated frequency-dependent anomalous refraction angle is plotted as the dashed curve in Fig. 3D, revealing excellent agreement with the experimental results. At 1.4 THz ($\lambda_0 = 214$ µm), the angle-dependent transmittance (Fig. 3E) shows a maximum output power of 61% at $\theta_t = 24°$ and reveals minimal transmission at negative angles (0.6% at –24°). Figure 3D also shows that the anomalous refraction intensity drops to zero as $\lambda_0$ approaches the periodicity $D_1$ corresponding to 0.54 THz, at which $\theta_t$ approaches 90°. This indicates that the incident wave is either totally reflected or converted to surface waves (*27*).

Our devices are ultrathin and operate within the technologically relevant THz frequency range, where many important functionalities, including polarization conversion, beam steering, and wavefront shaping, have been extremely challenging to accomplish. Our results have shown that we can achieve broadband high performance linear polarization conversion and near-perfect anomalous refraction. Additional numerical simulations show that near-perfect anomalous reflection can also be accomplished using the same concept (Fig. S8). Note that no particular optimization was undertaken to increase the conversion efficiency and bandwidth. Our demonstrations can be extended to other relevant frequencies. However, fabrication challenges

;

and metal losses can become issues when approaching visible frequencies that significantly degrade the device performance. Our results form the foundation for more advanced applications; for instance, an appropriately constructed device can serve as a high-performance spatial light modulator. The wavefront shaping can result in a helical phase dependence forming Laguerre-Gauss modes carrying an orbital angular momentum that can acquire any integer value (*28*), which is useful in quantum entanglement (*2929*) and enables opportunities in telecommunications (*30*).

**References and Notes:**

1. M. Born, E. Wolf, *Principles of Optics*. (Pergamon, ed. 6th, 1980).

2. L. Young, L. A. Robinson, C. A. Hacking, *IEEE Trans. Antennas Propag.* **21**, 376 (1973).

3. J. S. Tharp, B. A. Lail, B. A. Munk, G. D. Boreman, *IEEE Trans. Antennas Propag.* **55**, 2983 (2007).

4. D. R. Smith, W. J. Padilla, D. C. Vier, S. C. Nemat-Nasser, S. Schultz, *Phys. Rev. Lett.* **84**, 4184 (2000).

5. J. B. Pendry, *Phys. Rev. Lett.* **85**, 3966 (2000).

6. D. Schurig *et al.*, *Science* **314**, 977 (2006).

7. J. B. Pendry, A. J. Holden, D. J. Robbins, W. J. Stewart, *IEEE Trans. Microw. Theory Tech.* **47**, 2075 (1999).

8. J. M. Hao *et al.*, *Phys. Rev. Lett.* **99**, 063908 (2007).

9. J. Y. Chin, M. Z. Lu, T. J. Cui, *Appl. Phys. Lett.* **93**, 251903 (2008).

10. M. Euler, V. Fusco, R. Cahill, R. Dickie, *Iet Microw Antenna P* **4**, 1764 (2010).

11. Y. Q. Ye, S. He, *Appl. Phys. Lett.* **96**, 203501 (2010).

12. C. Menzel *et al.*, *Phys. Rev. Lett.* **104**, 253902 (2010).

13. N. I. Zheludev, E. Plum, V. A. Fedotov, *Appl. Phys. Lett.* **99**, 171915 (2011).

14. Z. Y. Wei, Y. Cao, Y. C. Fan, X. Yu, H. Q. Li, *Appl. Phys. Lett.* **99**, 221907 (2011).

15. W. J. Sun, Q. He, J. M. Hao, L. Zhou, *Opt. Lett.* **36**, 927 (2011).

16. M. Mutlu, A. E. Akosman, A. E. Serebryannikov, E. Ozbay, *Phys. Rev. Lett.* **108**, 213905 (2012).

17. J. K. Gansel *et al.*, *Science* **325**, 1513 (2009).

18. Y. Zhao, M. A. Belkin, A. Alu, *Nat. Commun.* **3**, 870 (2012).

19. X. G. Peralta *et al.*, *Opt. Express.* **17**, 773 (2009).

20. A. C. Strikwerda *et al.*, *Opt. Express.* **17**, 136 (2009).

21. Y. J. Chiang, T. J. Yen, *Appl. Phys. Lett.* **102**, 011129 (2013).

22. N. F. Yu *et al.*, *Science* **334**, 333 (2011).

23. X. J. Ni, N. K. Emani, A. V. Kildishev, A. Boltasseva, V. M. Shalaev, *Science* **335**, 427 (2012).

24. H.-T. Chen *et al.*, *Phys. Rev. Lett.* **105**, 073901 (2010).

25. H.-T. Chen, *Opt. Express* **20**, 7165 (2012).

26. See supplementary materials on *Science* Online.

27. S. L. Sun *et al.*, *Nat. Mater.* **11**, 426 (2012).

28. S. Franke-Arnold, L. Allen, M. Padgett, *Laser Photon. Rev.* **2**, 299 (2008).

29. R. Fickler *et al.*, *Science* **338**, 640 (2012).

30. J. Wang *et al.*, *Nat. Photon.* **6**, 488 (2012).

31. M. A. Ordal, R. J. Bell, R. W. Alexander, L. L. Long, M. R. Querry, *Appl. Opt.* **24**, 4493 (1985).

32. J. F. O'Hara, J. M. O. Zide, A. C. Gossard, A. J. Taylor, R. D. Averitt, *Appl. Phys. Lett.* **88**, 251119 (2006).

33. D. Grischkowsky, S. Keiding, M. Vanexter, C. Fattinger, *J. Opt. Soc. Am. B.* **7**, 2006 (1990).

34. Note, however, that non-zero-order Bragg modes are necessary to fully describe how electromagnetic energy is distributed within the metamaterial structure. See Y. Zeng, H.-T. Chen, and D. A. R. Dalvit, *Opt. Express* **21**, 3540 (2013).

**Acknowledgments:** We acknowledge partial support from the Los Alamos National Laboratory LDRD program. This work was performed, in part, at the Center for Integrated Nanotechnologies, a U.S. Department of Energy, Office of Basic Energy Sciences user facility. Los Alamos National Laboratory, an affirmative action equal opportunity employer, is operated by Los Alamos National Security, LLC, for the National Nuclear Security Administration of the U.S. Department of Energy under Contract No. DE-AC52-06NA25396.

**Supplementary Materials:**

Materials and Methods

Figures S1-S8

References (*31-34*)

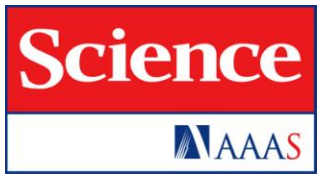

# Supplementary Materials for

## Terahertz Metamaterials for Linear Polarization Conversion and Anomalous Refraction

Nathaniel K. Grady, Jane E. Heyes, Dibakar Roy Chowdhury, Yong Zeng, Matthew T. Reiten, Abul K. Azad, Antoinette J. Taylor, Diego A. R. Dalvit, and Hou-Tong Chen

Correspondence to: chenht@lanl.gov

**This PDF file includes:**

Materials and Methods
Figs. S1 to S8
References (*31-34*)

**Materials and Methods**

Numerical Simulations

All of our numerical simulations were carried out using the commercial available numerical package CST Microwave Studio 2011. The simulations were performed by using its frequency domain solver with periodic boundary conditions. The metal structures, including the ground plane, the gratings, and the resonator array, were made of 200 nm thick gold films. In the simulations the gold was modeled as a Drude metal with plasma frequency $f_p$ = 2181 THz and collision frequency $f_c$ = 6.5 THz (*31*). The polyimide was simulated with a dielectric function of $\varepsilon = 3(1+0.05i)$. For all simulations supposed to have normal incidence, we add a small angle ($\theta_i = 0.5°$) to ensure that polarization modes are uniquely defined. The results of the simulations are the S-parameters including amplitude and phase. When simulating the reflection and transmission coefficients of an individual interface, we use a unit cell only containing the relevant components. For instance, Fig. S1 shows the resonant co- and cross-polarized reflection and transmission coefficient (amplitude and phase) for a simulation unit cell consisting of air, cut-wire, and polyimide.

Sample Fabrication

Photolithography, electron-beam metal deposition, and lift-off were used to fabricate the metallic structures. Spin coating and thermal curing were used to apply the polyimide layers (PI-2525). The sample shown in Fig. 1 for the linear polarization converter in reflection was fabricated by the following steps:

1) 10 nm / 200 nm thick Ti/Au layers were deposited on a bare GaAs substrate by using electron-beam evaporation.

2) A 33 µm thick polyimide layer was then spin coated, and thermally cured at 300°C for 150 minutes with a 2.5°C/min temperature ramp.

3) The cut-wire array was defined by standard photolithographic methods, deposition of 10 nm / 200 nm thick Ti/Au films, and a lift-off process.

The freestanding samples for the linear polarization converter in transmission depicted in Fig. 2 and anomalous refraction illustrated in Fig. 3 were fabricated in the following way:

1) A 4 µm thick polyimide layer was first spin coated on a bare GaAs wafer and subsequently cured at 300°C for 150 minutes.

2) Standard photolithographic methods, deposition of 10 nm / 200 nm thick Ti/Au films, and a lift-off process were used to fabricate the back grating.

3) A 33 µm (for the sample in Fig. 2) or 28 µm (for the sample in Fig. 3) thick polyimide layer was spin coated and thermally cured at 300°C for 150 minutes.

4) The resonator array was fabricated as in step 2.

5) A second 33 µm (for the sample in Fig. 2) or 28 µm (for the sample in Fig. 3) thick polyimide layer was deposited as in step 3.

6) The front metal grating was fabricated as in step 2.

7) A layer of 4 µm thick polyimide was spin coated and thermally cured as in step 1.

8) Finally, we mechanically removed the GaAs substrate by peeling off the polyimide encapsulated structure to establish the freestanding samples.

All of the fabricated samples are 1 cm × 1 cm except for the sample for anomalous refraction, which has an area of 2 cm × 2 cm.

Terahertz Time-Domain Spectroscopy for Experimental Characterization

We characterized the devices using a fiber-coupled photoconductive dipole antenna based terahertz time-domain spectrometer (THz-TDS) (*32, 33*) incorporating four wire-grid polarizers, which can be configured to operate either in reflection or transmission (see Fig. S4). Due to mechanical restrictions our minimal incidence angle is 25°. A mode-locked Ti:Sapphire oscillator provides near infrared femtosecond laser pulses with a central wavelength of 800 nm, pulse duration of ~50 fs, and repetition rate of 76 MHz. The laser pulses excite a photoconductive dipole antenna under a DC bias, which radiates single cycle broadband THz pulses. Two pairs of polyethylene lenses are used to collimate and focus the THz beam. The first pair of lenses (L1 and L2) collimates and then focuses the THz radiation to a spot with diameter of ~3 mm onto the sample. We use the second pair of lenses (L3 and L4) to collimate and focus the reflected or transmitted THz radiation onto the photoconductive dipole antenna detector. The collection optics and detector were mounted on an arm that can rotate around the sample. The signal is measured using a lock-in amplifier. The laser pulses are coupled to the THz detector using a single-mode optical fiber. This allows us to measure the reflection or transmission as a function of angle by rotating the THz detector without changing the time delay between the THz and the detecting near-infrared pulses. In order to maintain the short pulse duration at the detector, a pair of diffraction gratings is used to pre-compensate for fiber dispersion. For transmission measurements with normal incidence, the maximum measured transmission angle is ~70° in either direction. The usable frequency range of the system is about 0.2 – 2.5 THz.

In the measurements we recorded the electric field of the THz signal in the time-domain. We obtained the amplitude and phase spectra of the transmission or reflection by applying a Fast Fourier Transform (FFT) to the time domain data, which were then normalized to an appropriate reference spectrum. For transmission measurements, the reference was free-space (without the sample). For reflection, the reference was a gold-coated substrate identical to the one the sample was fabricated on.

Theoretical Formulation

An exact theoretical treatment of the electromagnetic scattering processes in our multilayered metamaterial structures requires expanding the fields into Bragg modes within each of the layers and taking into account the couplings between the several diffraction orders. Since the operational wavelength is much larger than the typical dimensions of the unit cell, a good approximation is to keep only zero-order Bragg modes when analyzing the scattering process, as inter-order couplings are expected to be weak (*34*). We confirm the validity of this approximation *a posteriori* after comparison with the numerical and experimental results for the far-field reflected and transmitted fields.

We use two different, but equivalent, methods to solve for the zero-order electric and magnetic fields within each of the layers. The first method is based on tracking the various Fabry-Pérot-like scattering processes within the structures. Fig. S5 shows an example of these scattering paths for the structure shown in Fig. 1. Each of the metallic structured layers in Figs. 1-3, i.e., gratings and cut-wire (resonator) array, are treated as zero thickness impedance-tuned surfaces, which modify the zero-order reflection and transmission coefficients at the cap-grating-spacer, spacer-cut-wire-spacer, and spacer-grating-substrate interfaces. These coefficients are obtained through numerical simulations. For the simulation of each interface, two orthogonal port modes are used at each port representing $x$- and $y$-polarized input and output plane waves, and therefore there are a total of 16 S-parameters corresponding to the reflection and transmission coefficients $r_{b\sigma';a\sigma}$ and $t_{b\sigma';a\sigma}$, where the subscripts indicate that light is incident from medium $a$ with polarization $\sigma(=x, y)$ and propagates to medium $b$ with polarization $\sigma'$. The second method we use is the transfer matrix approach, which is more convenient when the number of layers becomes large, as in Figs. 2 and 3. For a single interface with a planar metamaterial structure between two boundary media $a$ and $b$, the 4×4 transfer matrix $M_{ba}$ of the interface relates forward and backward propagating fields on each side of the interface:

$$\begin{pmatrix} E_{bx}^{(f)} \\ E_{by}^{(f)} \\ E_{bx}^{(b)} \\ E_{by}^{(b)} \end{pmatrix} = M_{ba} \begin{pmatrix} E_{ax}^{(f)} \\ E_{ay}^{(f)} \\ E_{ax}^{(b)} \\ E_{ay}^{(b)} \end{pmatrix}. \quad (1)$$

The transfer matrix can be expressed in terms of the reflection and transmission coefficients as follows

$$M_{ba} = \begin{pmatrix} 1 & 0 & -r_{bx;bx} & -r_{bx;by} \\ 0 & 1 & -r_{by;bx} & -r_{by;by} \\ 0 & 0 & t_{ax;bx} & t_{ax;by} \\ 0 & 0 & t_{ay;bx} & t_{ay;by} \end{pmatrix}^{-1} \begin{pmatrix} t_{bx;ax} & t_{bx;ay} & 0 & 0 \\ t_{by;ax} & t_{by;ay} & 0 & 0 \\ -r_{ax;ax} & -r_{ax;ay} & 1 & 0 \\ -r_{ay;ax} & -r_{ay;ay} & 0 & 1 \end{pmatrix}. \quad (2)$$

Here the superscripts ($f$) and ($b$) indicate forward and backward propagating light, the subscripts $x$ and $y$ indicate the polarization state of the field in the medium ($a$, $b$), and $r$ and $t$ are the previously computed reflection and transmission coefficients. The 4×4 propagation matrix within a given homogeneous medium $a$ (thickness $d_a$, refractive index $n_a(\omega)$) is given by $P_a = \text{diag}(e^{ik_0 n_a d_a}, e^{ik_0 n_a d_a}, e^{-ik_0 n_a d_a}, e^{-ik_0 n_a d_a})$, where $k_0$ is the free-space wave number. For a structure composed of several metamaterial interfaces surrounded by homogenous media, the overall transfer matrix is given by $M = \ldots M_{43} P_3 M_{32} P_2 M_{21}$.

For the metamaterial linear polarization converter in reflection (Fig. 1), the reflected field can be expanded as $E_{r,\sigma} = \sum_{j=1}^{\infty} E_{r,\sigma j}$, where $j - 1$ denotes the number of roundtrips within the spacer layer. The first few terms are given by $E_{r,\sigma j=1} = E_{i,x} r_{1\sigma;1x}$, $E_{r,\sigma j=2} = E_{i,x} \sum_{\sigma'} t_{1\sigma;2\sigma'} t_{2\sigma';1x} e^{2ik_0 nd}$, etc. These different terms, both for cross-polarized ($\sigma =$ *y*) and co-polarized ($\sigma = x$) reflections, are depicted in Figs. 1E, F. As seen from those figures, we obtain highly efficient cross-polarization conversion with the destructive interference between the co-polarized terms.

We now present an argument based on the Green's function of the structure shown in Fig. 1 to provide additional understanding of the mechanism behind the metamaterial polarization converter. The vector wave equation for the electric field is

$$\nabla \times \nabla \times \vec{E}(\vec{r},\omega) - \frac{\omega^2}{c^2}\varepsilon_a(\vec{r},\omega)\vec{E}(\vec{r},\omega) = i\omega\mu_0 \vec{J}_{CW}(\vec{r},\omega) + i\omega\mu_0 \vec{J}_i(\vec{r},\omega), \qquad (3)$$

where $\varepsilon_a(\vec{r},\omega)$ is the permittivity of an auxiliary system which is identical to the original structure except that the cut-wire layer has been replaced by a vacuum layer, the current $\vec{J}_{CW}(\vec{r},\omega) = i\omega\varepsilon_0\chi_{CW}(\vec{r},\omega)\vec{E}(\vec{r},\omega)$ is the polarization current inside the cut-wires ($\chi_{CW}$ is the susceptibility of the cut-wires), and the current $\vec{J}_i(\vec{r},\omega)$ is the source at infinity that generates the incident field. Using the scattering Green tensor $\underline{\underline{G}}_a(\vec{r},\vec{r}';\omega)$ of the differential operator in the left-hand-side of the above equation, the solution for the reflected field is therefore

$$\vec{E}_r(\vec{r},\omega) = i\omega\mu_0\int d\vec{r}'\underline{\underline{G}}_a(\vec{r},\vec{r}';\omega)\bullet\vec{J}_i(\vec{r}',\omega) + i\omega\mu_0\int d\vec{r}'\underline{\underline{G}}_a(\vec{r},\vec{r}';\omega)\bullet\vec{J}_{CW}(\vec{r}',\omega). \qquad (4)$$

We now consider the experimental setup in which the incident field is *x*-polarized. Due to the isotropy of $\varepsilon_a(\vec{r},\omega)$ and $\underline{\underline{G}}_a(\vec{r},\vec{r}';\omega)$, the auxiliary system does not modify the polarization of the incident field. Hence, the first term in Eq. (4) can be simply written as $R(\omega)E_{i,x}(\vec{r},\omega)\hat{x}$, where $R(\omega)$ is the Fresnel reflection coefficient of the auxiliary system for light impinging from the vacuum side. On the other hand, the incident field generates electric currents $\vec{J}_{CW}$ inside the cut-wires. Since the cut-wires are oriented at α = 45° with respect to the incident field, $\vec{J}_{CW}$ has nearly identical *x* and *y* components. The second term in Eq. (4) then takes the form $i\omega\mu_0\int d\vec{r}'G_{a,xx}(\vec{r},\vec{r}';\omega)J_{CW,x}(\vec{r}',\omega)\hat{x} + i\omega\mu_0\int d\vec{r}'G_{a,yy}(\vec{r},\vec{r}';\omega)J_{CW,y}(\vec{r}',\omega)\hat{y}$. We see that it is possible to reduce, and even nullify, the co-polarized reflected field when $R(\omega)E_{i,x}(\vec{r},\omega)$ destructively interferes with $i\omega\mu_0\int d\vec{r}'G_{a,xx}(\vec{r},\vec{r}';\omega)J_{CW,x}(\vec{r}',\omega)$.

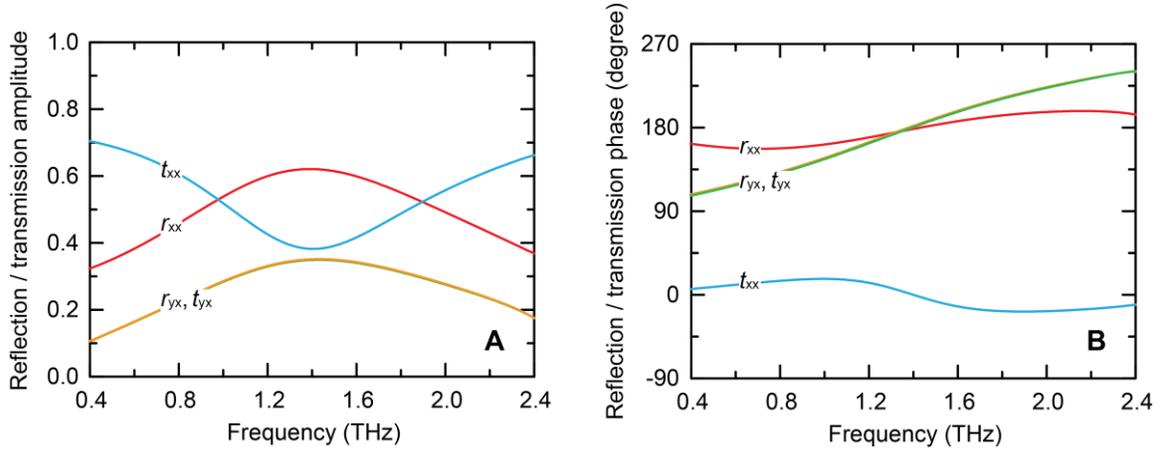

**Fig. S1.** Co- and cross-polarized reflection and transmission amplitude (**A**) and phase (**B**) spectra for light impinging from vacuum onto a single layer of cut-wire array on polyimide substrate. In the reflection $r_{\sigma'\sigma}$ and transmission $t_{\sigma'\sigma}$ coefficients, $\sigma$ and $\sigma'$ indicate the input and output polarizations (*x* or *y* direction), respectively.

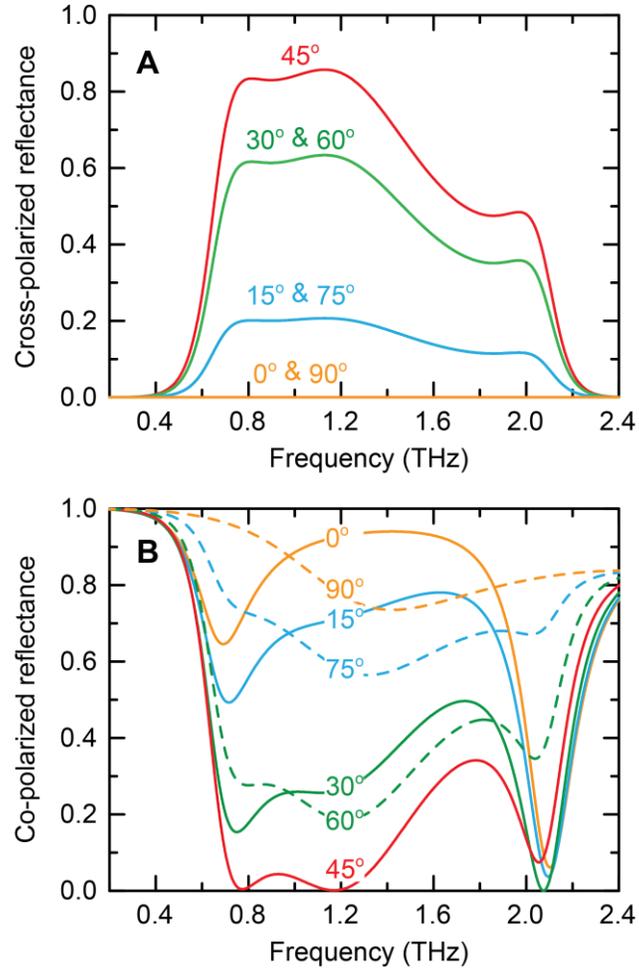

**Fig. S2.** (**A**) Cross- and (**B**) co-polarized reflectance for the metamaterial polarization converter shown in Fig. 1A for different angles $\alpha$ between the input polarization and the cut-wire long axis.

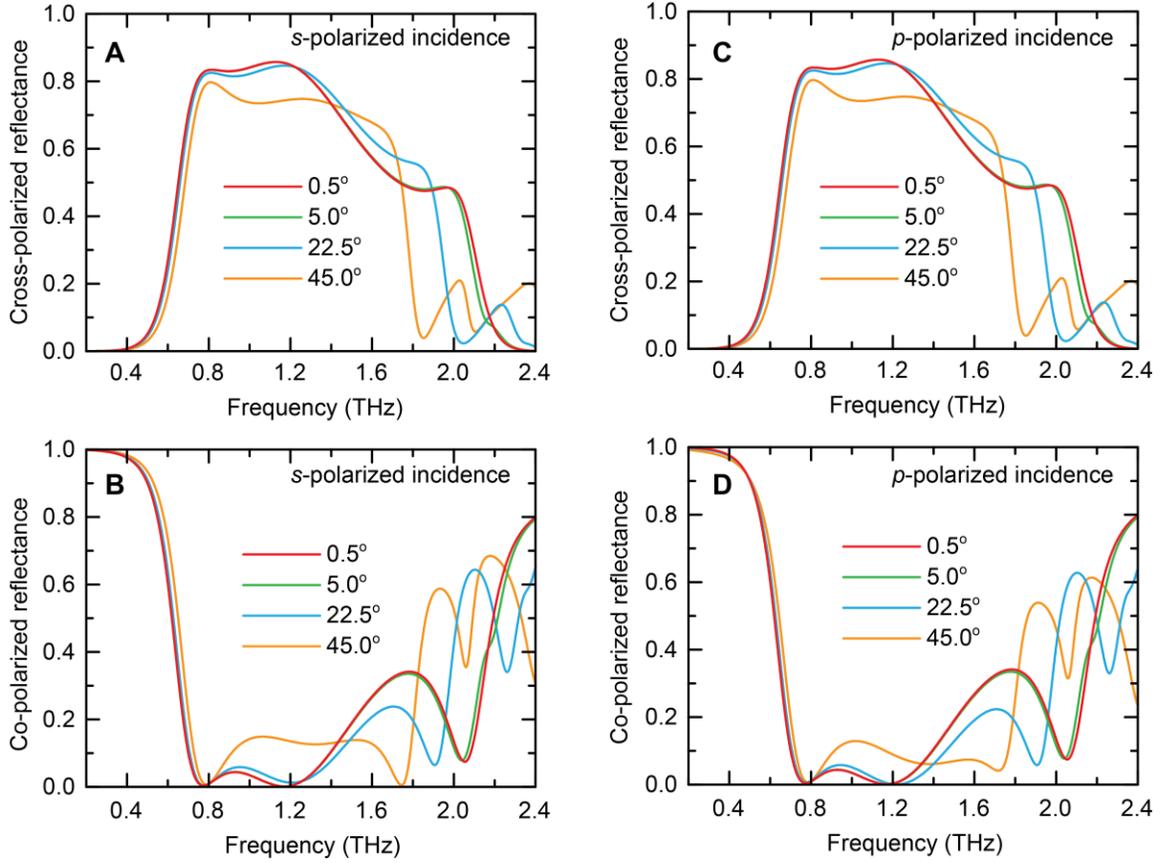

**Fig. S3.** Numerical simulated results on incidence angle $\theta_i$ dependence of cross- and co-polarized reflectance for s-polarized (**A**, **B**) and p-polarized (**C**, **D**) incidence.

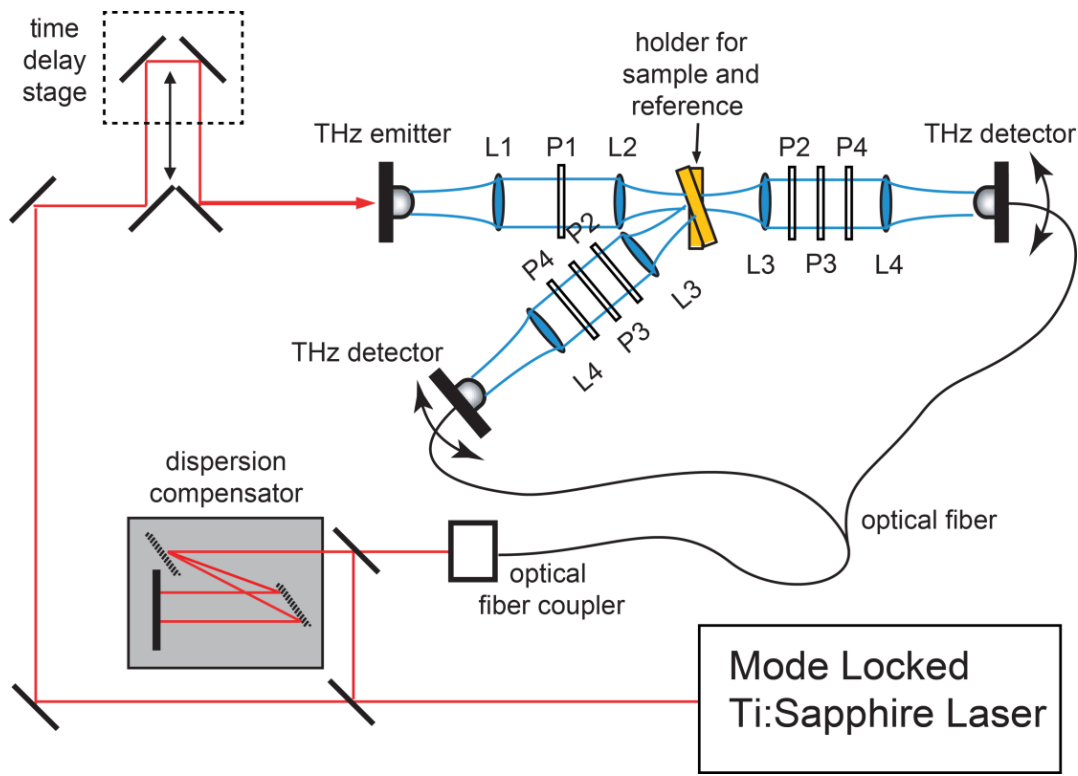

**Fig. S4.** Schematic of the fiber-coupled photoconductive dipole antenna based THz time-domain spectrometer. The system can be configured to measure either reflection or transmission by rotating the detector arm around the sample. The sample holder can also be rotated, which enables the measurement of input and output angular dependent reflection/transmission. P1-P4 are wire-grid polarizers: P1 is aligned with the incident polarization, P2 is either parallel (co-polarization) or perpendicular (cross-polarization) to P1, P3 is 45° with respect to P2, and P4 is parallel to P1 (and to the polarization of the THz detector).

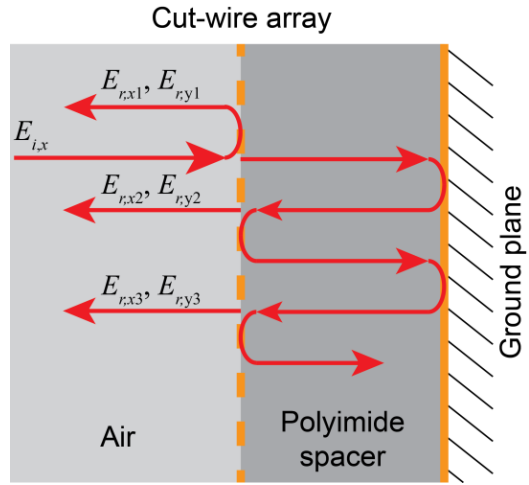

**Fig. S5.** Schematic of multiple reflections in the Fabry-Pérot-like metamaterial polarization converter in reflection.

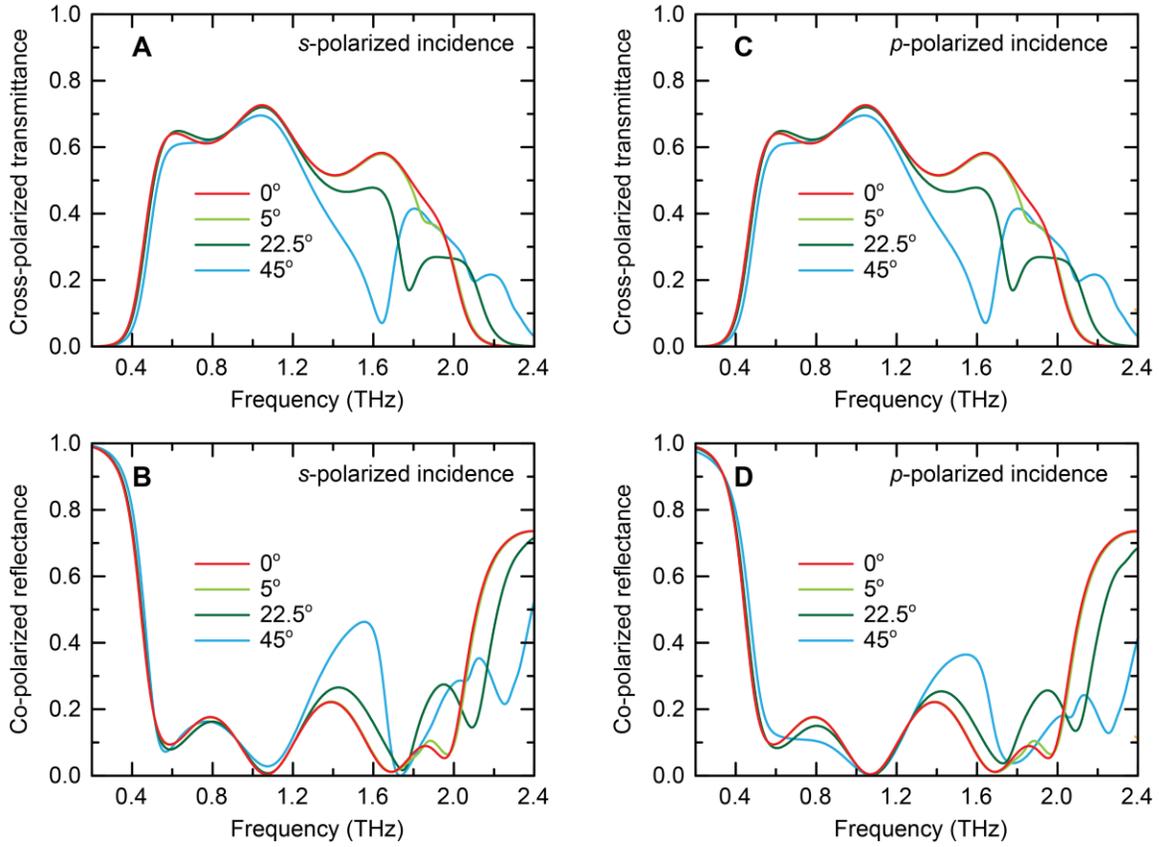

**Fig. S6.** Numerical simulated results on incidence angle $\theta_i$ dependent cross-polarized transmittance for *s*- and *p*-polarized incidence (**A**, **C**), and co-polarized reflectance for *s*- and *p*-polarized incidence (**B**, **D**).

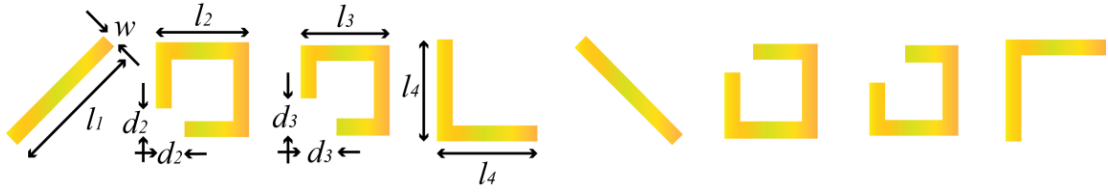

**Fig. S7.** Structural dimensions for the resonators used in the super-unit-cell shown in Fig. 3A for the demonstration of metamaterial anomalous refraction: $w = 8$ μm, $l_1 = 68$ μm, $l_2 = 46$ μm, $l_3 = 44$ μm, $l_4 = 50$ μm, $d_2 = 9$ μm, $d_3 = 18$ μm, and the super-unit-cell outer dimensions are $D_1 = 560$ μm and $D_2 = 70$ μm. The four resonators at the right side are just flipped along the horizontal axis of the four resonators at the left side. The same super-unit-cell with a few slightly different parameters ($d_2 = 14$ μm, $D_1 = 500$ μm) was also used to replace the single resonant element in Fig. 1A for the numerical demonstration of near-perfect anomalous reflection shown in Fig. S8.

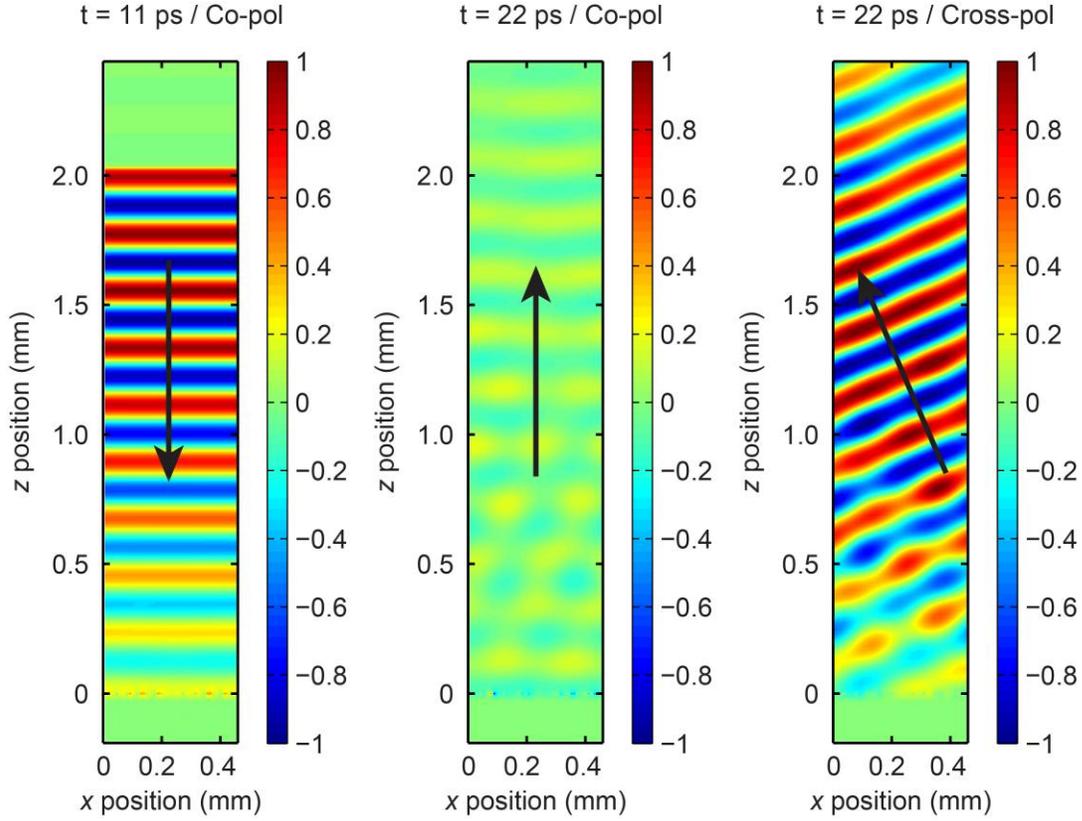

**Fig. S8.** Metamaterial near-perfect anomalous reflection demonstrated through numerical simulations. The color maps the electric field of the normally incident quasi-monochromatic Gaussian beam (left panel), co-polarized ordinary reflection (center panel), and cross-polarized anomalous reflection (right panel). The arrows indicate the wave propagation direction. The metamaterial is located at $z = 0$. The values of $t$ indicate the time after the beginning of the THz signal generation at $z = 2$ mm. The metamaterial structure is based on the one shown in Fig. 1A by replacing the single resonator with the super-unit-cell shown in Fig. S7. The spacer thickness is 28 µm, and the operational frequency is 1.35 THz.

**References and Notes:**

1. M. Born, E. Wolf, *Principles of Optics*. (Pergamon, ed. 6th, 1980).
2. L. Young, L. A. Robinson, C. A. Hacking, *IEEE Trans. Antennas Propag.* **21**, 376 (1973).
3. J. S. Tharp, B. A. Lail, B. A. Munk, G. D. Boreman, *IEEE Trans. Antennas Propag.* **55**, 2983 (2007).
4. D. R. Smith, W. J. Padilla, D. C. Vier, S. C. Nemat-Nasser, S. Schultz, *Phys. Rev. Lett.* **84**, 4184 (2000).
5. J. B. Pendry, *Phys. Rev. Lett.* **85**, 3966 (2000).
6. D. Schurig *et al.*, *Science* **314**, 977 (2006).
7. J. B. Pendry, A. J. Holden, D. J. Robbins, W. J. Stewart, *IEEE Trans. Microw. Theory Tech.* **47**, 2075 (1999).
8. J. M. Hao *et al.*, *Phys. Rev. Lett.* **99**, 063908 (2007).
9. J. Y. Chin, M. Z. Lu, T. J. Cui, *Appl. Phys. Lett.* **93**, 251903 (2008).
10. M. Euler, V. Fusco, R. Cahill, R. Dickie, *IET Microw. Antenna Propag.* **4**, 1764 (2010).
11. Y. Q. Ye, S. He, *Appl. Phys. Lett.* **96**, 203501 (2010).
12. C. Menzel *et al.*, *Phys. Rev. Lett.* **104**, 253902 (2010).
13. N. I. Zheludev, E. Plum, V. A. Fedotov, *Appl. Phys. Lett.* **99**, 171915 (2011).
14. Z. Y. Wei, Y. Cao, Y. C. Fan, X. Yu, H. Q. Li, *Appl. Phys. Lett.* **99**, 221907 (2011).
15. W. J. Sun, Q. He, J. M. Hao, L. Zhou, *Opt. Lett.* **36**, 927 (2011).
16. M. Mutlu, A. E. Akosman, A. E. Serebryannikov, E. Ozbay, *Phys. Rev. Lett.* **108**, 213905 (2012).
17. J. K. Gansel *et al.*, *Science* **325**, 1513 (2009).
18. Y. Zhao, M. A. Belkin, A. Alu, *Nat. Commun.* **3**, 870 (2012).
19. X. G. Peralta *et al.*, *Opt. Express.* **17**, 773 (2009).
20. A. C. Strikwerda *et al.*, *Opt. Express.* **17**, 136 (2009).
21. Y. J. Chiang, T. J. Yen, *Appl. Phys. Lett.* **102**, 011129 (2013).
22. N. F. Yu *et al.*, *Science* **334**, 333 (2011).
23. X. J. Ni, N. K. Emani, A. V. Kildishev, A. Boltasseva, V. M. Shalaev, *Science* **335**, 427 (2012).
24. H.-T. Chen *et al.*, *Phys. Rev. Lett.* **105**, 073901 (2010).
25. H.-T. Chen, *Opt. Express.* **20**, 7165 (2012).
26. See supplementary materials on *Science* Online.
27. S. L. Sun *et al.*, *Nat. Mater.* **11**, 426 (2012).
28. S. Franke-Arnold, L. Allen, M. Padgett, *Laser Photon. Rev.* **2**, 299 (2008).
29. R. Fickler *et al.*, *Science* **338**, 640 (2012).
30. J. Wang *et al.*, *Nat. Photon.* **6**, 488 (2012).
31. M. A. Ordal, R. J. Bell, R. W. Alexander, L. L. Long, M. R. Querry, *Appl. Opt.* **24**, 4493 (1985).
32. J. F. O'Hara, J. M. O. Zide, A. C. Gossard, A. J. Taylor, R. D. Averitt, *Appl. Phys. Lett.* **88**, 251119 (2006).
33. D. Grischkowsky, S. Keiding, M. Vanexter, C. Fattinger, *J. Opt. Soc. Am. B.* **7**, 2006 (1990).

34. Note, however, that non-zero-order Bragg modes are necessary to fully describe how electromagnetic energy is distributed within the metamaterial structure. See Y. Zeng, H.-T. Chen, and D. A. R. Dalvit, *Opt. Express* **21**, 3540 (2013).